\documentclass[structabstract]{aa}  
\usepackage{graphicx}
\usepackage{lscape}
\usepackage{txfonts}
\usepackage{natbib}
\usepackage{sidecap}
\usepackage{comment}
\usepackage{color}
\usepackage{booktabs}

\begin{document}
\title{Masgomas-4: Physical characterization of a double-core \\ obscured cluster with a massive and very young stellar population.}
\author{S. Ram\'irez Alegr\'ia\inst{1,2} \and A. Mar\'in-Franch\inst{3} \and A. Herrero\inst{4,5}}

\institute{The Millennium Institute of Astrophysics (MAS), Santiago, Chile.
\and Instituto de Astrof\'isica, Universidad  de Valpara\'iso, Valpara\'iso, Chile. \email{sebastian.ramirez@uv.cl} 
\and Centro de Estudios de F\'isica del Cosmos de Arag\'on (CEFCA), E-44001, Teruel, Spain. \email{amarin@cefca.es} 
\and Instituto de Astrof\'isica de Canarias, 38205 La Laguna, Tenerife, Spain. \email{ahd@iac.es} 
\and Departamento de Astrof\'isica, Universidad de La Laguna, E-38205 La Laguna, Tenerife, Spain. \\}
\date{Received 2013 / Accepted:}

\abstract
{The discovery of new, obscured massive star clusters has changed our understanding of the Milky Way star-forming activity from a passive to
a very active star-forming machine. The search for these obscured clusters is strongly supported by the use of all-sky, near-IR surveys.}
{The main goal of the MASGOMAS project is to search for and study unknown, young, and massive star clusters in the
Milky Way, using near-IR data. Here we try to determine the main physical parameters (distance, size, total mass, and age)
of Masgomas-4, a new double-core obscured cluster.}
{Using near-IR photometry ($J$, $H$, and $K_S$) we selected a total of 21 stars as OB-type star candidates. Multi-object, near-IR follow-up
 spectroscopy allowed us to carry out the spectral classification of the OB-type candidates.}
{Of the 21 spectroscopically observed stars, ten are classified as OB-type stars, eight as 
F- to early G-type dwarf stars, and three as late-type giant stars. Spectroscopically estimated distances indicate 
that the OB-type stars belong to the same cluster, located at a distance of $1.90^{+1.28}_{-0.90}$ kpc. 
Our spectrophotometric data confirm a very young and massive stellar population, with a clear concentration 
of pre-main-sequence massive candidates (Herbig Ae/Be) around one of the cluster cores. The presence of a surrounding
H\,II cloud and the Herbig Ae/Be candidates indicate an upper age limit of 5 Myr.}
{}
 \titlerunning{Masgomas-4}
\maketitle


\section{Introduction}\label{msgms4}

 Massive stars affect the interstellar medium at several scales (by ionizing their surrounding media, 
depleting their native clouds through their own birth, or modifying the star formation rate in their 
neighbourhood), and are fundamental pieces in Galactic evolution \citep{martins05}. Their influence 
on the Galaxy takes place during their very short periods, compared with less massive stars lifetimes. 
We often find them deeply embedded in their natal obscured massive clusters.

 Owing to high-extinction in the Galactic plane, the detection of distant Galactic massive stars and clusters using optical
data becomes impossible. Near-infrared photometric surveys such as 2MASS \citep{skrutskie06}, GLIMPSE 
\citep{benjamin03}, UKIDSS \citep{lawrence07}, and the ESO public survey VISTA Variables in the V\'ia L\'actea (VVV, 
 \citealt{minnitiVVV10,saito12}) have allowed the discovery of  candidates in the Galaxy: for example 
 \citet{bdb03,bica03}, \citet{dutra03}, and \citet{froebrich07} using 2MASS photometry; \citet{davies11}, \citet{kurtev08}, 
 \citet{kurtev07}, and \citet{mercer05} from GLIMPSE data; and \citet{borissova11} from Vista-VVV data. 

Nevertheless, the census of massive clusters is far from complete; up to 100 clusters with masses greater than $10^4$ 
$M_{\odot}$ may remain hidden \citep{hansonpopescu08}. Systematic search programmes for these objects are necessary 
for a full understanding of our Galaxy. This has motivated us to develop the MASGOMAS project \citep{marin09}. 
The present phase of the project is a systematic search for massive cluster candidates in the Galactic disc.
Using the preliminary version of the search algorithm we found two new massive cluster candidates  
in the direction of the Scutum--Centaurus arm base. This very interesting region in the Milky Way hosts a 
concentration of massive clusters with a red supergiant population: RSGC1 \citep{figer06}, RSGC2 \citep{davies07}, RSGC3
 \citep{alexander09,clark09}, Alicante\,7 \citep{negueruela11}, Alicante\,8, \citep{negueruela10}, and Alicante\,10 \citep{gonzalezfernandez12}. 
Our first cluster discovered in this direction, Masgomas-1 \citep{ramirezalegria12}, is a spectroscopically confirmed massive cluster with an initial 
total mass of $(1.94\pm0.28)\cdot10^4 {M}_{\odot}$ and a coexisting population of OB-type and RSG stars.

\begin{figure*}
\centering
\includegraphics[width=15.5cm]{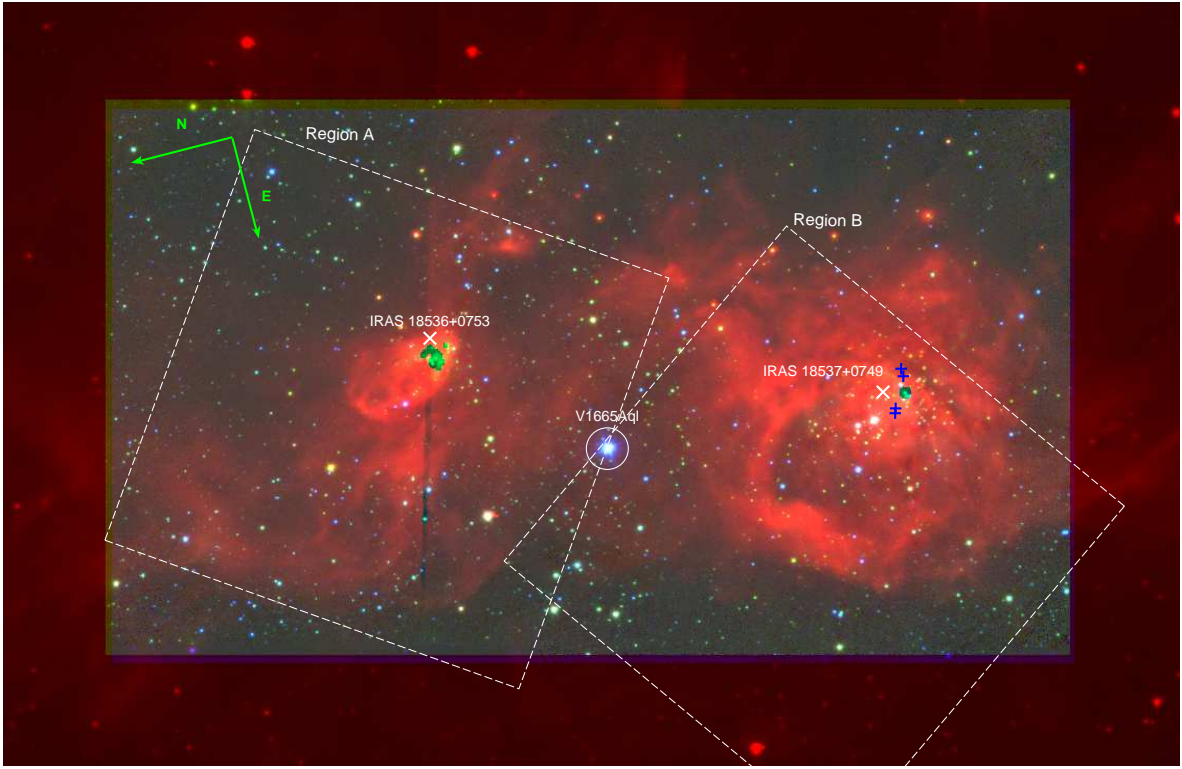}
       \caption{Masgomas-4 false colour image ($J$=blue, $K_S$=green, $5.8\,\mu m$=red). $J$ and $K_S$ images are from 
       LIRIS, and the $5.8\,\mu m$ image is from IRAC, Spitzer. The size of Spitzer image is $12'\times7'$, and the size of the false colour 
       image is $9.2'\times5.2'$. White crosses show the central positions of the IRAS sources and blue crosses show the position of the 
       masers (water, ammonia, and methanol) reported in Section \ref{msgms4}. The variable object \object{V 1665 Aql} is marked 
       with a white circle. The arrows indicate the image orientation and are 1 arcminute in length. White squares show the orientation 
       of both multi-object masks (masks A and B).}
       \label{m04_color}
\end{figure*}

A second cluster candidate, Masgomas-4, is located in the coordinates $l=40.530^{\circ}$, $b=+2.576^{\circ}$, or 
${\alpha}_{2000}=18^{\mathrm {h}}56^{\mathrm {m}}07^{\mathrm {s}}$, ${\delta}_{2000}=+07^{\circ}56'07''$. 
The candidate is found embedded a bright mid-IR nebulosity, as shown in Figure \ref{m04_color}. Two
IRAS sources are located close to Masgomas-4: IRAS 18536+0753 and IRAS 18537+0749. Their positions are 
also presented in Figure \ref{m04_color}.
 
 Based on these IRAS sources, we define two regions for Masgomas-4: region A (centred around IRAS 18536+0753) and
region B (centred around IRAS 18537+0749). These regions have been extensively investigated
in radio wavelength, but there are no studies of their stellar population\footnote{An optical spectrophotometric study by
\citet{forbes89} included five stars from the Masgomas-4 field. Spectral classification was reported only for one B7\,V star, but the author 
does not associate this star with the H\,II region.}. In region A we can find methanol \citep{slysh99,szymczak00} and hydroxyl 
\citep{baudry97} masers, both indicative of possible massive star formation. No distance estimates are available
in the literature for region A.

In region B we find the H\,II region Sh2-76 E \citep{sharpless59,zinchenko97}, which can also be associated with methanol 
and ammonia masers. For Sh2-76 E there are two distance estimates: one of 2.1 kpc, obtained by \citet{plume92} using 
water maser observations, and a second estimate of 2.2 kpc by \citet{valtts00}, using methanol maser observations ([HHG86] 
185345.9+074916). The source Sh2-76 is also linked with the previously mentioned IRAS source IRAS 18537+0749. \citet{chan96,chan95Cat}
report it as a massive young stellar object and cite two distance estimates: 2.1 kpc \citep{zinchenko94} and 2.33 kpc \citep{palagi93}.

 In spite of the evidence of massive star formation in both regions A and B, there are no works dedicated to the observation and 
 characterization of their stellar population. There is only a spectral classification of B7--9\,V \citep{ibanoglu07} for \object{V 1665 Aql}, 
 an eclipsing binary located between regions A and B. This binary system can be identified as a bright and blue source, with $K_S=7.7$ 
 mag and $(J-K_S)\sim0$ mag, and according to \citet{ibanoglu07}, it is located at a distance of 477 pc. We will show 
 later that this implies that the eclipsing binary is a foreground object. 
 
 Because there are no previous works dedicated to its stellar population, we cannot state whether Masgomas-4 is a single
 cluster or both regions A and B are two different objects. We expect to solve this question by using the individual distance and 
 extinction estimates. In either case, we will also carry out a complete physical characterization (mass, age, star formation activity) 
 of both regions.
 
In this article we describe the near-IR observations of Masgomas-4 (Section \ref{observations}) and the near-IR photometric 
 diagrams and spectral classifications of the 21 spectroscopically observed stars (Section \ref{resultados}). In Section \ref{discusion_m04} 
 we complete an analysis of the physical parameters (distance for individual regions, cluster distance, mass, and age), and we try to determine 
 whether Masgomas-4 is a single cluster or two different objects. We summarize our results in Section \ref{conclusiones}.

 \begin{figure*}
\centering
\includegraphics[width=15.0cm,angle=0]{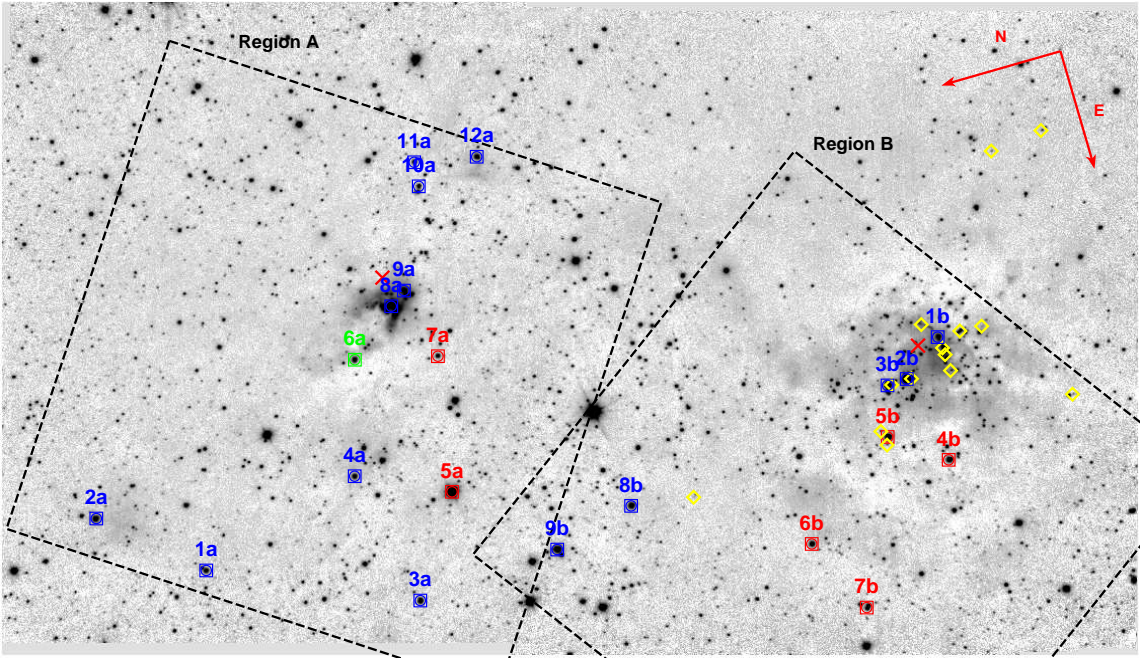}
       \caption{Masgomas-4 $K_S$ image obtained with LIRIS. Stars with spectral classification are marked
       with blue (OB-type dwarfs), green (A, F, and G dwarfs), and red squares (giant stars). We also mark with 
       yellow rhombi the Herbig Ae/Be candidates and with red crosses the position of the IRAS sources. 
       Orientation is the same as in Figure \ref{m04_color} to help comparison between the figures. Length 
       of orientation arrows is 1 arcmin, and black squares show the orientation of both multi-object masks (regions A and B).}
       \label{m04_ks}
\end{figure*}
 

\section{Observations}
\label{observations}

We use data obtained with the LIRIS instrument \citep{manchado04}, a near-infrared imager/spectrograph mounted 
at the Cassegrain focus of the 4.2 m William Herschel Telescope (La Palma, Canary Island, Spain). The camera is equipped with a
 Hawaii 1024$\times$1024 HgCdTe array detector, with a field of view of $4.2'\times4.2'$ and a spatial scale of $0\,\farcs25\mathrm{~pixel}^{-1}$.
 This work includes broadband near-infrared imaging using filters $J$ (${\lambda}_C=1{.}250~\mu m$, $\Delta \lambda=0{.}160~\mu m$), $H$ 
 (${\lambda}_C=1{.}635~\mu m$, $\Delta\lambda=0{.}290~\mu m$), and ${K}_{S}$ (${\lambda}_C=2{.}150~\mu m$, $\Delta\lambda=0{.}320~\mu m$), 
 and multi-object mid-resolution $R\sim2500$ near-infrared spectroscopy, using pseudogrisms $H$ and $K$ \citep{fragoso2008}. 

The candidate apparent size is larger than the LIRIS field of view. Therefore, we divide it into two regions. Region A,
centred around ${\alpha}_{2000}=18^{\mathrm {h}}56^{\mathrm {m}}05.95^{\mathrm {s}}$, ${\delta}_{2000}=+07^{\circ}56'52.6''$
and including the IRAS source IRAS 18536+0753, and region B, centred on ${\alpha}_{2000}=18^{\mathrm {h}}56^{\mathrm 
{m}}11.09^{\mathrm {s}}$, ${\delta}_{2000}=+07^{\circ}53'06.5''$ and containing the IRAS source IRAS 18536+0749. The 
angular distance between both centres is 4\arcmin16\arcsec.
 
\begin{table*}
\caption{Summary of imaging and spectroscopic observations}

\begin{center}
\begin{tabular}{cccccc}
\toprule
Observing mode & Date & Filter & Exp. time & Seeing & Airmass\\
                                  &           &           &        [s]        &   [$''$] &    \\
\midrule
Masgomas-4 field imaging        	& 2010 June 23 	&  $J$   	& 108.0	&  0.78   &  1.17--1.21 \\
(A, B, and control)                    	& 2010 June 23 	&  $H$  	&  79.2      &  0.72  &  1.22--1.27  \\
   						& 2010 June 23 	& $K_S$	&  79.2    	&  0.70   &  1.28--1.36  \\
\midrule
MOS (R$\sim$2500)  	& 2011 September 16       & $H$       &  2520.0       &  0.83   &  1.18--1.35\\
Mask region A	         	& 2011 September 16       & $K$	&  1920.0	     &  0.86   &   1.40--1.63 \\
\midrule
MOS (R$\sim$2500)  	& 2011 September 14	& $H$       &  1800.0       &  0.78  &   1.47--1.75 \\
Mask region B			& 2011 September 14	& $K$	&  2880.0	     &  1.59   &   1.07--1.10 \\
\midrule
Telluric standard (R$\sim$2500)  	& 2011 Sept 14	& $H$       &  1080.0   & 0.95  &  1.59--1.94 \\
			                                  & 2011 Sept 14	& $K$	&  1800.0   & 1.12  &  1.07--1.19  \\
                                                           & 2011 Sept 16	& $H$       &  1040.0   & 1.16  &  1.50--1.88  \\
			                                  & 2011 Sept 16	& $K$	&    960.0   & 1.63   &  1.94--2.56  \\
\bottomrule
\end{tabular}
\end{center}
\label{msgms04_tabla_obs}
\end{table*}

  We obtained the cluster and control field images on 2010 June 23, with seeing values between $0.70-0.80''$
(see Table \ref{msgms04_tabla_obs}). We observed using a nine-point dithering mode to improve 
cosmic-ray reduction and bad-pixel reduction, and to construct the sky image for subtraction at a later time. For image reduction (bad-pixel 
mask, flat correction, sky subtraction, and alignment), we followed a similar procedure to the one described 
by \citet{ramirezalegria12}. Data reduction was made with FATBOY \citep{eikenberry06} and the LIRIS reduction 
package LIRISDR\footnote{http://www.iac.es/project/LIRIS}. The final $K_S$ image shown in Figure \ref{m04_ks} is a composite of regions A and B 
individual images.
 
 Instrumental photometry was made with DAOPHOT II, ALLSTAR, and ALLFRAME \citep{stetson94}. It was cleaned of 
 non-stellar and poorly measured objects using the sharp index (-0.25 $<$ sharp $<$ 0.25) and PSF fitting $\sigma < 0.1$. For 
 photometric calibration, we used 37 stars from region A and 48 stars from region B. These 
stars, part of the 2MASS catalogue \citep{skrutskie06}, are isolated and non-saturated stars in our photometry. For 
stars brighter than  $J<14$ mag, $H<11.5$ mag, or $K_S<9.5$ mag, we assumed the magnitudes from the 2MASS 
catalogue. 

\begin{figure}
\centering
\includegraphics[width=6.5cm,angle=270]{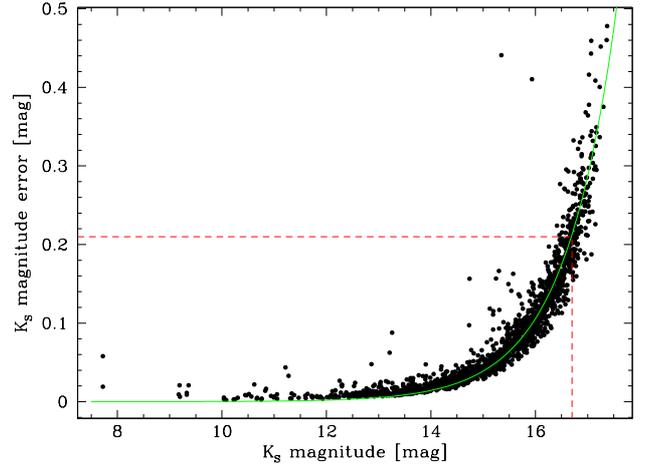}
       \caption{Calibrated $K_S$ magnitude versus photometric error for Masgomas-4. We show with a green solid line the exponential function
       fitted to the data, and with a red segmented line the limiting magnitude of $K_S\sim$16.7 mag associated with the photometric error of 0.21 
       mag (or a signal-to-noise ratio of 5).}
       \label{limit_Ks}
\end{figure}

For our data we define the limiting magnitude as the characteristic magnitude with a photometric error of 0.21 mag. This photometric error 
limit corresponds to a signal-to-noise ratio of 5 \citep{newberry91}. In Figure \ref{limit_Ks}, we show the $K_S$ magnitude and its associated error for the 
Masgomas-4 photometry, and the exponential function fitted to the data. Using this function we estimated that the limiting magnitude for
the $K_S$ band is 16.7 mag.

We used SKYCAT for the astrometric calibration, by correlating image and equatorial coordinates for the same set of stars used in the 
photometric calibration. Fitting errors are less than 0.15 arcsec for the three filters.

 Near-infrared spectra were observed during 2011 September 14, 15, and 16 (half nights, Table \ref{msgms04_tabla_obs}). 
 As shown in next section, two masks were designed to observe the OB-type star candidates. We included
 in the region A mask (hereafter mask A) twelve stars, and in the region B mask (mask B), nine stars. To avoid large signal-to-noise 
 differences for spectra with the same exposure times, we limited the magnitude difference in each mask to less than 2 mag 
 (for $K_S$). Slits of 0.8 arcsec in width and 8-10 arcsec in length were cut on each mask, which produced a resolution of R~2500. 
 The typical values for the signal-to-noise is $\sim76.7$ for $H$ band spectra, and $\sim78.6$ for $K$ band spectra, 
 per resolution element. We also took into account the slit position along the dispersion axis, a key factor to obtain the right 
 wavelength range to observe OB-type characteristic spectral lines. The telluric standard for all spectra was the A0\,V star \object{HD\,231033}, 
 and the modelling of the A0\,V spectral contribution was carried out with XTELLCOR \citep{vacca03}. The subtraction of the telluric 
 lines and the correction due to the airmass difference were completed with the IRAF task TELLURIC.
 


\section{Results}\label{resultados}

In the first part of this section we show the photometric diagrams (colour--magnitude, colour--colour, and pseudocolour--magnitude) 
for regions A, B, and the control field. The second part of the section is dedicated to the infrared spectra from Masgomas-4, and 
the description of the spectral features used for the spectral classification.

\subsection{Photometry}\label{diagramasM04}

\begin{figure*}
\centering
\includegraphics[width=13.0cm,angle=270]{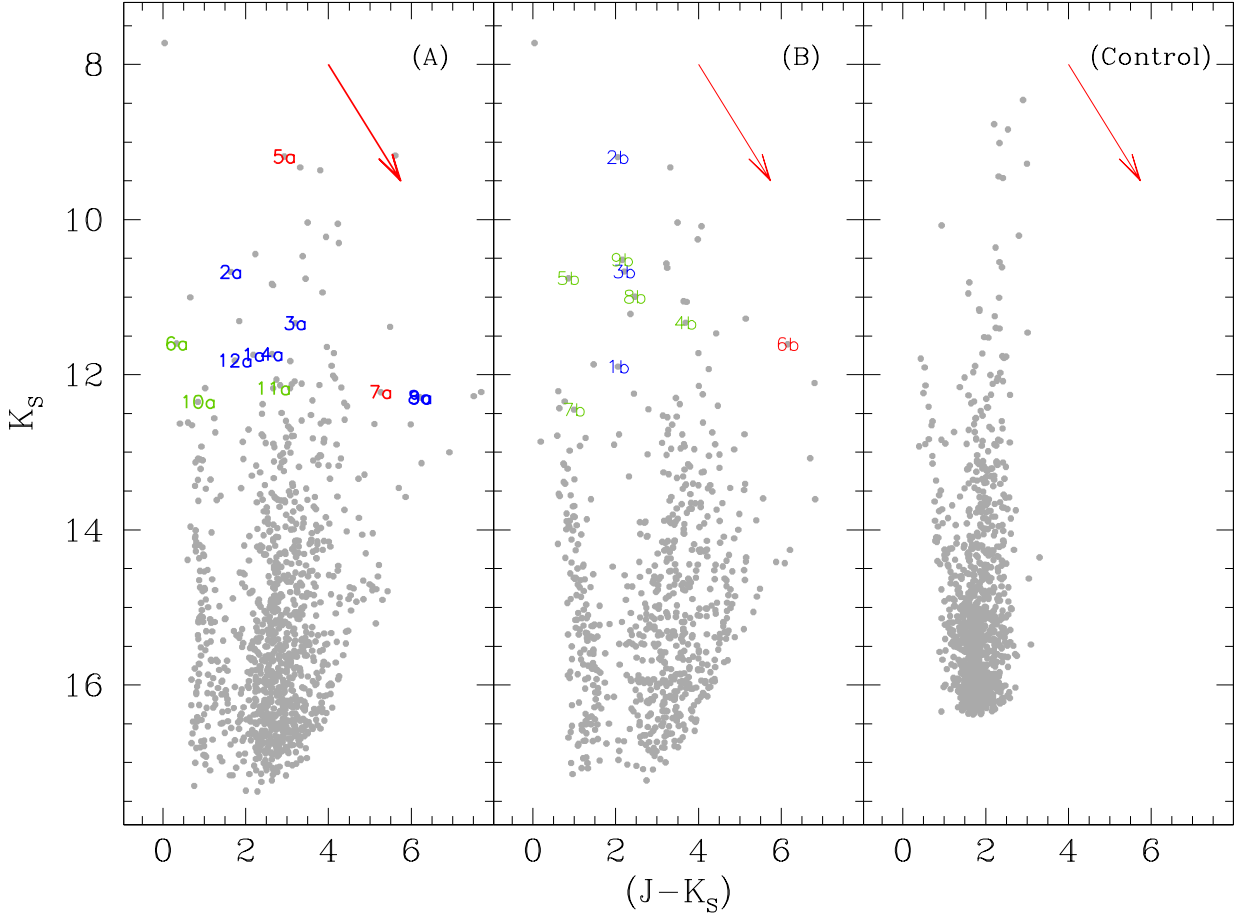}
       \caption{Cluster colour--magnitude diagrams (CMD) for regions A (left), B (middle), and the control field (right). 
       We label the spectroscopically observed and classified stars using their numbering for masks A and B. 
       We indicate the OB-type stars with blue symbols; A-, F-, and G-type dwarfs with green symbols; and giant stars
       with red symbols. The red arrow shows an extinction of $A_{K_S}$=1.5 mag, assuming the extinction law from \citet{rieke89}, with 
       $R=3.09$ \citep{rieke85}.}
       \label{m04_cmd}
\end{figure*}

In the colour--magnitude diagrams (CMD) for regions A and B (Figure \ref{m04_cmd}) the cluster main sequence is not 
clearly seen. Differential extinction spread it to redder colours, following the reddening vector. This makes it difficult to distinguish the
cluster stars among the field stars using only the CMD. It is therefore necessary to use extra photometric colours to clearly spot the cluster 
stellar population. For this purpose we used the reddening-free parameter $Q_{IR}$ \citep{comeron05, negueruela07} to identify the OB-type
star candidates in regions A and B. 

\begin{figure*}
\centering
\includegraphics[width=7.5cm,angle=270]{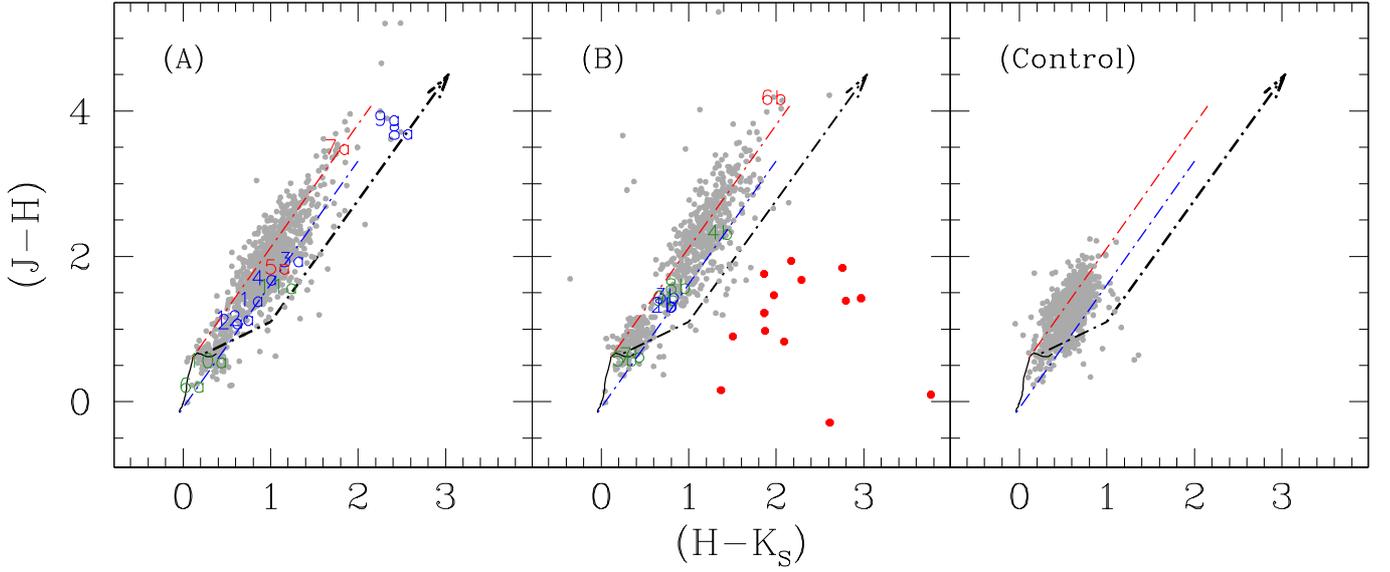}
       \caption{Cluster CCDs for regions A (left), B (middle), and the control field (right).
       We use the same symbols as in Figure \ref{m04_cmd} to mark the spectroscopically observed stars. In the 
       diagrams we mark with segmented lines the position of an O8\,V (blue lines) and a K5\,V star (red lines), following 
       the Rieke extinction law \citep{rieke89}, as used in this work. Thicker black segmented lines show the expected 
       position for T Tauri stars along the extinction vector. In the region B diagram, we show with red circles the stars 
       with infrared excess, candidates to Herbig Ae/Be objects.}
       \label{m04_ccd}
\end{figure*}

The definition of the parameter $Q_{IR}$ depends on the assumed extinction law, which can be determined from the regions'  
colour--colour diagrams (CCD, Figure \ref{m04_ccd}). These diagrams show that the extinction in regions A and B is accurately described by the Rieke 
extinction law \citep{rieke89}, with $R=3.09$ \citep{rieke85}. 

\begin{figure*}
\centering
\includegraphics[width=13.0cm,angle=270]{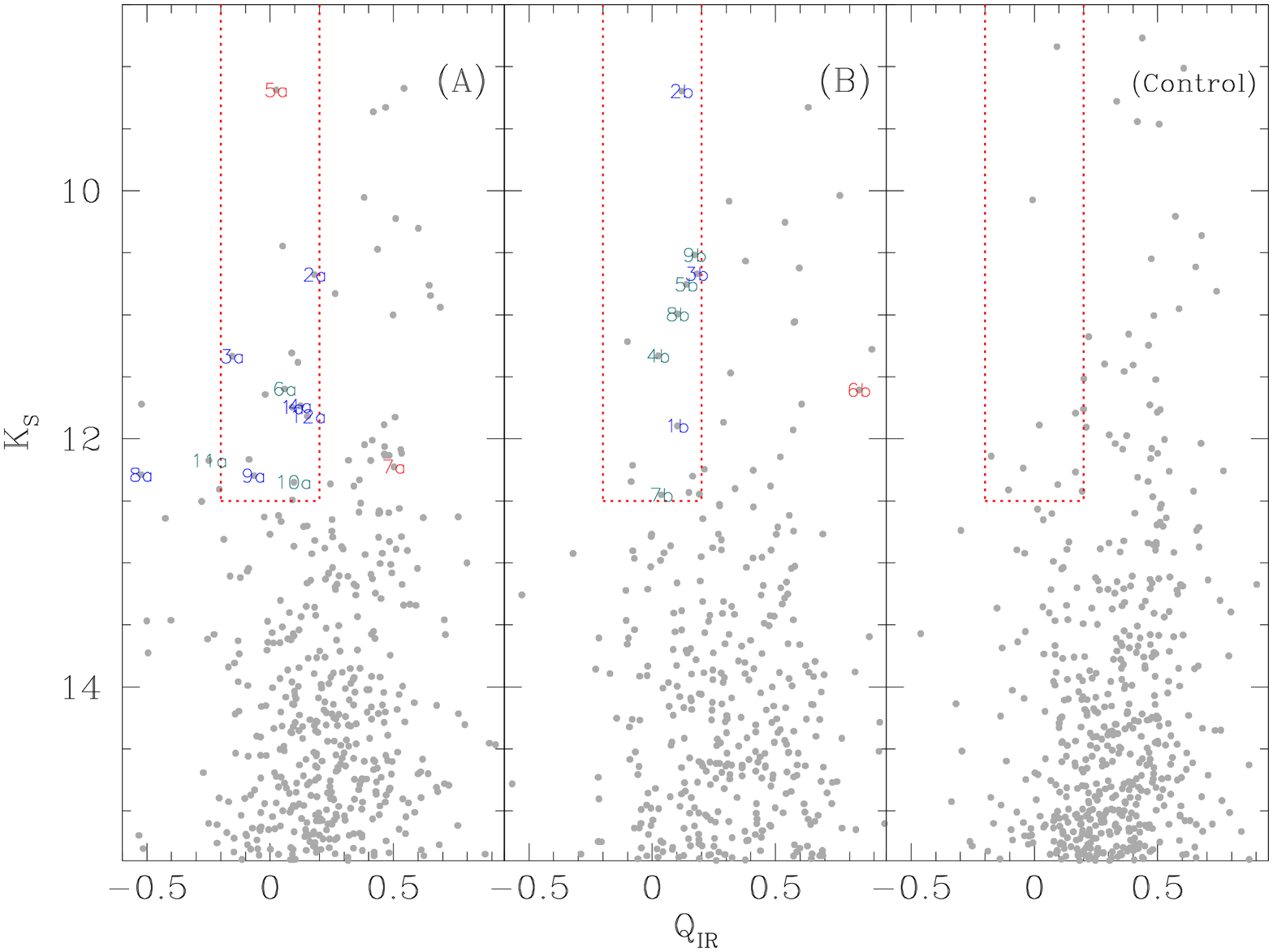}
       \caption{Cluster pseudocolour--magnitude diagrams for regions A (left), B (middle), and the control 
       field (right). We use the same symbols as in Figure \ref{m04_cmd} to mark the 
       spectroscopically observed stars (blue=OB-type, green=A-, F-, G-type, red=giants). Dotted 
       red lines mark the adopted limits in pseudocolour $Q_{IR}$ and magnitude $K_S$ for the OB-candidates 
       selection.}
       \label{m04_qmd}
\end{figure*}

The slopes of the linear function fitted to the $(H-K_S)$-$(J-H)$ diagrams are

\begin{equation}
\frac{\mathrm{E(J-H)}}{\mathrm{E(H-K_S)}} = 1.69\pm0.03
\end{equation}
for region A and 

\begin{equation}
\frac{\mathrm{E(J-H)}}{\mathrm{E(H-K_S)}} = 1.66\pm0.04
\end{equation}
for region B. These values are in  agreement, within errors, with the value derived from the Rieke extinction law:

\begin{equation}
\frac{\mathrm{E(J-H)}}{\mathrm{E(H-K_S)}} = 1.70.
\end{equation}
Using the reddening-free parameter $Q_{IR}$ defined by this extinction law

\begin{equation}
Q_{IR} = (J-H) - 1.7\cdot(H-K_S),
\end{equation}
we can see in both pseudocolour--magnitude diagrams a vertical sequence around $Q_{IR}\sim0$ mag (Figure \ref{m04_qmd}). 
These sequences, which are understood as OB-type star candidates, are not seen in the control field's 
pseudocolour--magnitude diagram. Therefore, it is a feature associated with the cluster stellar population.

  In the colour--magnitude diagrams we can also see a group of bright and reddened 
stars, close to $K_S\sim10$ mag and $(J-K_S)\sim4$ mag. These stars could be part of the upper end of
the cluster main sequence, spread by the differential extinction to dimmer magnitudes and redder colours. Because 
these objects have pseudocolour values of $Q_{IR}\sim0.6$ mag, they are most probably giant stars from the galactic disc.
 
 For multi-object spectroscopy we selected stars with reddening-free parameter $Q_{IR}$ between 
$-0.2$ mag and 0.2 mag. We also selected some stars, outside this $Q_{IR}$ range, due to their position in the 
Masgomas-4 field (see Fig. \ref{m04_ks}). For example, star 7a ($Q_{IR}=0.50$ mag) is located at the centre of region 
A; star 8a has $Q_{IR}=-0.52$ mag and is also a central star in region A; star 6b ($Q_{IR}=0.84$ mag) is located on the edge 
of a bubble-shaped structure in region B; star 11a has a value of $Q_{IR}=0.25$ mag and, because it does not overlap 
with any other slit in the mask, we decided to include it.

A second group of stars are blue objects with an interesting position in the cluster field. For example star 6a, 
which is located in one arm-shaped structure of IRAS 18536+0753, and star 5b, which is close to 
the bubble-shaped structure of IRAS 18537+0749. 

 Finally we identify a group of very reddened stars in the region B CCD. This 
could indicate the existence of a younger (and more embedded) population in region B. These objects are 
shown with yellow rhombi in Figure \ref{m04_ks}.

\subsection{Spectral classification}\label{clasif_m04}

 Spectral classification for Masgomas-4 has followed a similar procedure to that in previous group works (G61.48+0.09 \citealt{marin09}; 
 Sh2-152, NGC\,7538 \citealt{puga10}; \citealt{ramirezalegria11}; and Masgomas-1, \citealt{ramirezalegria12}). The classification scheme is based on 
 the detection of absorption lines and the comparison of their depth and shape with similar resolution spectra 
 of known spectral types. The catalogues used for spectral classification were \citet{hanson96} ($K$ band) and \citet{hanson98} 
 ($H$ band), for OB-type stars, and \citet{meyer98,wallacehinkle97} for late-type stars. For the assigned spectral type 
 we assume an error of $\pm 2$ subtypes, similar to \citet{hanson10}, \citet{negueruela10}, and \citet{ramirezalegria12}.

\begin{figure*}
\centering
\includegraphics[width=15.0cm]{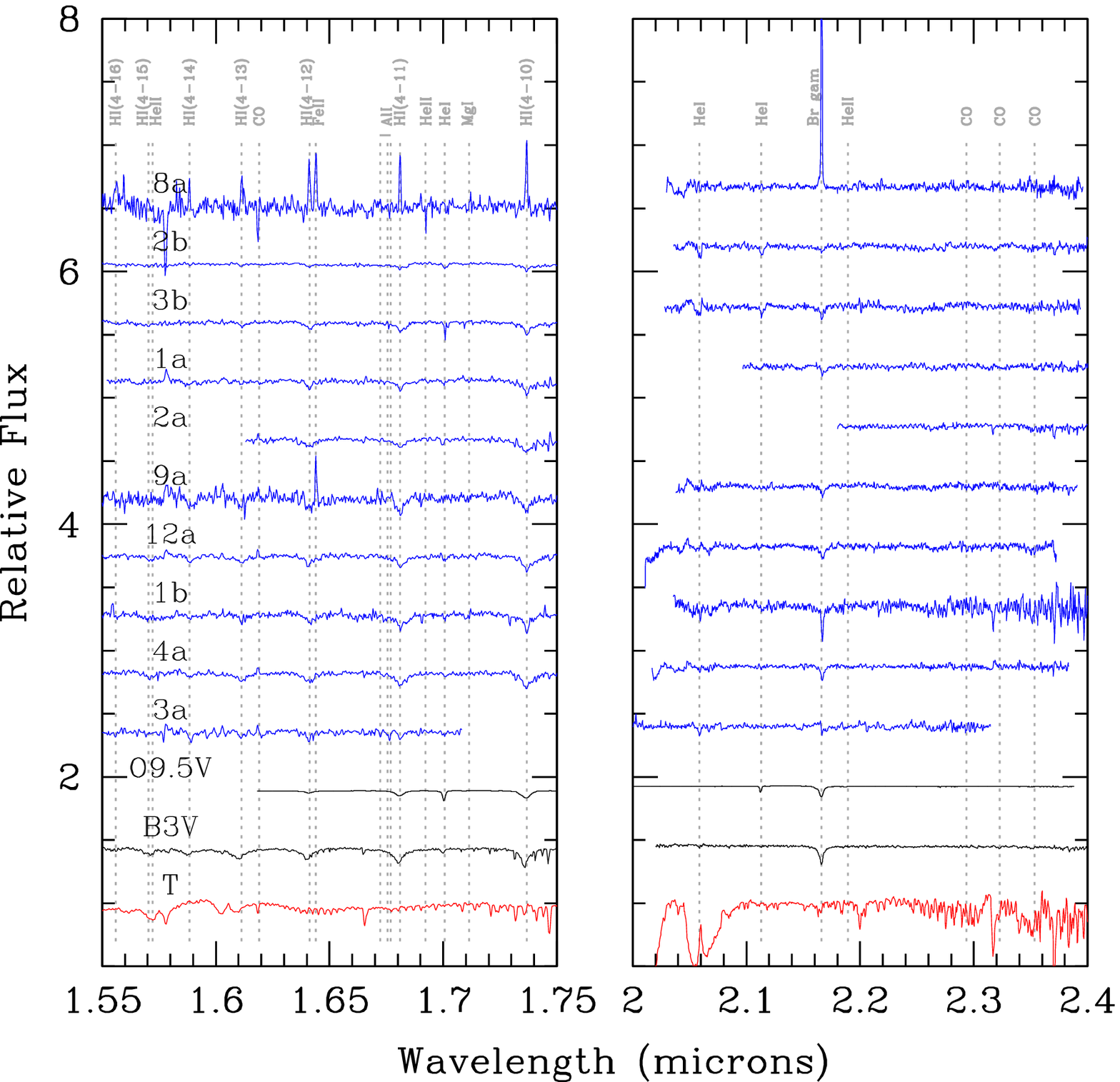}
       \caption{Individual spectra in $H$ (left) and $K$ band (right) for OB-type stars. Spectral features used in
       the spectral classification are labelled in grey. We show in black two spectra from the spectral libraries for comparison:
       O9.5\,V HD\,37468 (\citealt{hanson05}, degraded to our resolution) and B3\,V HR5191 \citep{meyer98}. The telluric correction spectrum is shown in red.}
       \label{m04_espectrosA}
\end{figure*}

\begin{figure*}
\centering
\includegraphics[width=15.0cm]{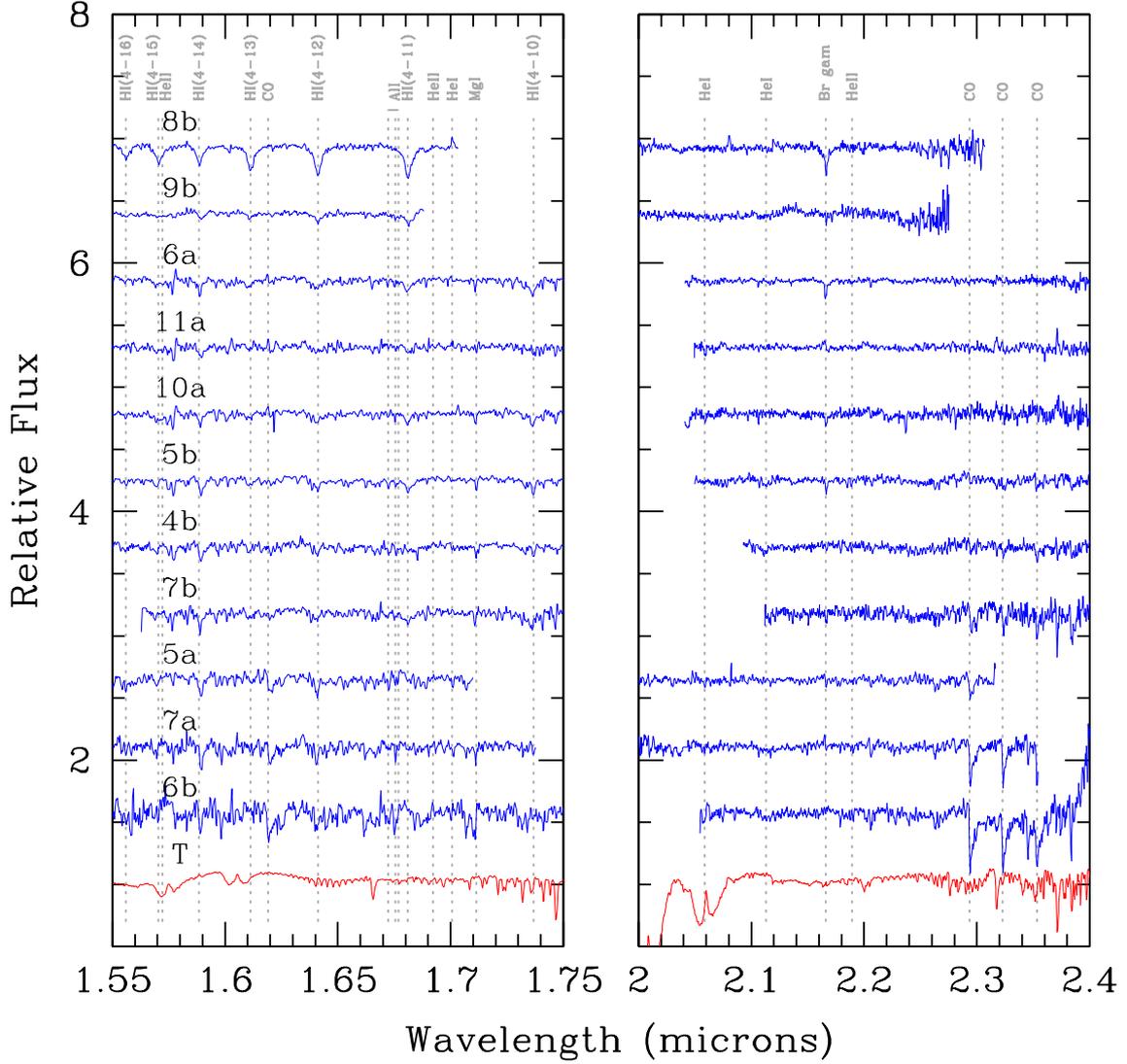}
       \caption{Individual spectra in $H$ (left) and $K$ band (right) for the AFG-type dwarfs and the giant stars. We show in grey the 
       spectral lines used for the spectral classification. The red spectrum corresponds to the telluric correction spectrum, shown for spectral 
       feature comparison.}
       \label{m04_espectrosB}
\end{figure*}

 Table \ref{msgms04_data_stars} provides the coordinates, infrared magnitudes, and spectral types of the  
spectroscopically observed stars. The mask A and mask B final spectra, as well as the telluric correction spectra, are 
shown in Figures \ref{m04_espectrosA} (OB-type stars) and \ref{m04_espectrosB} (AFG-type dwarfs 
and giants stars).

 \begin{table*}
\caption{Stars with observed spectra. We include the equatorial coordinates, infrared magnitudes ($J$, $H$, $K_S$), 
and spectral type for each star. For those stars with estimated luminosity class, we have included the extinction and 
distance estimates.}
\begin{center}
\scalebox{0.9}{
\begin{tabular}{ccccccccc}
\toprule
 ID & RA (J2000)                                                & dec (J2000)                                 & $J$     & $H$   & $K_S$ & Spectral type & $A_{K}$ & Distance \\
      & [$\mathrm{{~~}^{h}{~~}^{m}{~~}^{s}}$] & [$\mathrm{{~~}^{\circ}~~'~~''}$] & [mag] & [mag] & [mag]   &                            & [mag]      & [kpc]          \\
 \midrule
 \multicolumn{9}{l}{OB-type dwarfs:} \\
 \midrule
 2b  & 18 56 11.98 & +07 53 38.1  &  $11.246\pm0.008$  &  $9.911\pm0.008$    & $9.196\pm0.006$    &  O9--9.5\,V     &	1.47$^{+0.02}_{-0.03}$ & 1.77$^{+0.26}_{-0.67}$ \\ 
 3b  & 18 56 11.99 & +07 53 48.1  &  $12.875\pm0.003$  &  $11.419\pm0.004$  & $10.767\pm0.004$  &  B1\,V             &	1.54$^{+0.03}_{-0.02}$ & 2.08$^{+1.43}_{-0.99}$ \\ 
 1a  & 18 56 11.05 & +07 59 31.0  &  $13.926\pm0.014$  &  $12.519\pm0.010$  & $11.744\pm0.006$ &  B2--3\,V       & 	1.52$^{+0.03}_{-0.02}$ & 2.47$^{+2.24}_{-0.77}$  \\
 2a  & 18 56 08.31 & +08 00 14.3  &  $12.318\pm0.018$  &  $11.217\pm0.002$  & $10.676\pm0.005$ &  B2--3\,V       & 	1.16$\pm$0.01 & 1.52$^{+0.95}_{-0.29}$  \\  
 9a  & 18 56 04.21 & +07 57 17.9  &  $18.526\pm0.014$  &  $14.628\pm0.046$  & $12.296\pm0.020$ &  B2--3\,V        &	4.19$^{+0.03}_{-0.02}$ & 0.93$^{+0.84}_{-0.29}$  \\ 
12a & 18 56 00.73 & +07 56 24.3  &  $13.554\pm0.028$  &  $12.404\pm0.005$  & $11.818\pm0.007$ &  B3\,V             &	1.21$\pm0.02$                & 2.13$^{+1.93}_{-0.22}$  \\ 
 1b  & 18 56 10.97 & +07 53 17.6  &  $13.954\pm0.005$  &  $12.620\pm0.005$  & $11.895\pm0.006$ &  B3\,V             &	1.43$\pm0.02$                & 2.00$^{+1.81}_{-0.21}$  \\ 
 4a  & 18 56 09.55 & +07 58 08.3  &  $14.370\pm0.007$  &  $12.665\pm0.007$  & $11.734\pm0.005$ &  B3--5\,V        & 	1.80$\pm$0.02                & 1.48$^{+0.67}_{-0.22}$  \\ 
 3a  & 18 56 14.12 & +07 57 56.5  &  $14.530\pm0.007$  &  $12.575\pm0.004$  & $11.335\pm0.011$ &  B0--5\,V       &	2.19$\pm$0.03                & 1.68$^{+1.19}_{-0.70}$  \\ 
 8a  & 18 56 04.57 & +07 57 26.1  &  $18.489\pm0.092$  &  $14.779\pm0.033$  & $12.289\pm0.021$ &  YSO	        & --                                           & --    \\   
 \midrule
 \multicolumn{9}{l}{A, F, G dwarfs:} \\
 \midrule
 8b  & 18 56 13.22 & +07 56 04.6  &  $13.459\pm0.003$  & $11.847\pm0.003$ & $11.100\pm0.005$ &  A5\,III--V       & 	1.58$^{+0.02}_{-0.03}$ & 0.38$^{+0.04}_{-0.03}$\\    
 9b  & 18 56 13.86 & +07 56 45.3  &  $12.678\pm0.005$  & $11.254\pm0.005$ & $10.615\pm0.010$ &  F3--7\,V        &	1.25$^{+0.03}_{-0.02}$ & 0.21$\pm0.00$\\   
 6a   & 18 56 05.90 & +07 57 50.9  & $11.926\pm0.021$ & $11.697\pm0.004$ & $11.597\pm0.008$ &  F6\,V             &	0.04$^{+0.03}_{-0.02}$ & 0.61$\pm0.01$  \\   
 11a & 18 56 00.29 & +07 56 53.9  & $14.839\pm0.009$ & $13.253\pm0.008$ & $12.175\pm0.006$ &  F6--8\,V        &	1.56$\pm0.02$                & 0.40$\pm0.04$  \\   
 10a & 18 56 01.09 & +07 56 55.4  & $13.195\pm0.031$ & $12.627\pm0.006$ & $12.349\pm0.008$ &  F7\,V             &	0.36$\pm0.02$                & 0.74$^{+0.08}_{-0.07}$  \\   
 5b  & 18 56 13.62 & +07 53 55.6   & $11.613\pm0.002$ & $11.022\pm0.002$ & $10.810\pm0.027$  &  F6--G1\,V     &	0.35$\pm0.02$                & 0.33$^{+0.05}_{-0.04}$  \\   
 4b  & 18 56 14.93 & +07 53 30.7   & $14.950\pm0.005$ & $12.661\pm0.006$ & $11.483\pm0.004$ &  G2--3\,V       &	2.19$^{+0.01}_{-0.02}$ & 0.15$^{+0.16}_{-0.01}$  \\   
 7b  & 18 56 18.78 & +07 54 30.9   &  $13.445\pm0.005$ & $12.788\pm0.011$ & $12.508\pm0.009$ &  G8--K0\,V     &	0.25$^{+0.03}_{-0.05}$ & 5.44$^{-0.54}_{+1.30}$  \\   
 \midrule
 \multicolumn{9}{l}{Giant stars:} \\
 \midrule
 5a  & 18 56 11.00 & +07 57 25.6   &  $12.121\pm0.018$ & $10.265\pm0.013$ & $9.188\pm0.021$  &  G4--7\,III        &	1.58$\pm0.02$               & 0.57$^{-0.03}_{+0.04}$  \\  
 7a  & 18 56 06.61 & +07 57 12.0   &  $17.490\pm0.036$ & $13.989\pm0.010$ & $12.225\pm0.005$ &  K0--2\,III       &	3.02$^{+0.05}_{-0.08}$ & 1.67$^{-0.32}_{+0.83}$  \\    
 6b  & 18 56 16.22 & +07 54 46.8   &  $17.764\pm0.051$ & $13.576\pm0.010$  & $11.606\pm0.004$  &  M0--1\,III   &	3.39$^{+0.01}_{-0.05}$ & 3.11$^{-0.27}_{+1.18}$   \\   
\bottomrule
\end{tabular}}
\end{center}
\label{msgms04_data_stars}
\end{table*}

According to the stellar spectra, we can separate the observed stars in three groups: OB-type stars, AFG-type dwarfs, and 
giant stars.

\subsubsection{OB-type stars}

These stars are identified by their helium lines and incomplete Brackett series. According to the observed spectral features, 
star 2b is the earliest object in our set. The spectrum has strong He\,I lines at 1.70 and 2.11\,$\mu m$, similar to those 
found in O9.5\,V stars (for example, HD\,37468 or HD\,149757, \citealt{hanson05}) and Brackett series similar to an O9\,V star (v.g. HD\,193322, 
\citealt{hanson96}). We have adopted a spectral type O9--9.5\,V for this star.

 Star 3b shows features of a slightly later star. The Brackett series is deeper and similar to the B2\,V HD\,36166 star \citep{hanson05}, 
and its clear He\,I lines are similar to the B0\,V HD\,36512 star \citep{hanson96}. We classify star 3b as B1\,V.

 Spectra of stars 1a, 2a, and 9a show weak Brackett series, until H\,I (4-12), with a depth similar to a B3\,V star (v.g.  HR\,5191, 
\citealt{meyer98}). The weak He\,I line at 1.70\,$\mu m$ indicates a stellar type similar to a B2\,V star, so we adopt a 
spectral type B2--3\,V for these stars. The spectrum of star 1a shows an emission at 1.58\,$\mu m$ and others stars (v.g. stars 3a, 6a, 9a, 10a, 
11a, and 12a) show an emission bump at this position. Inspecting the telluric correction spectrum, we see an absorption 
at the same position. We understand it to be an artifact from the reduction process and not a real stellar line. The Brackett series for 
star 2a is wider than the other star series, indicating that star 2a could be a fast rotator.

In the case of spectrum 9a, we also see an emission feature close to the H\,I (4-12) line, also seen in the 8a spectrum, that can 
be identified as an Fe\,II line. This line indicates the presence of circumstellar material, and can be found in Herbig Be stars
\citep{puga10}. Because for star 9a we observe its Brackett series in absorption, we understand that this object is as very young, 
but already in the main sequence. 

 In the spectra of stars 12a and 1b we observe an $H$ band spectrum similar to a B3\,V star \citep{meyer98}, with a Brackett 
 series extending until HI\,(4-15), and a clear He\,I at 1.70\,$\mu m$. The $H$ band spectrum indicates a B3\,V spectral type 
 for both stars.

 The Brackett series for spectrum 4a extends until H\,I (4-15) line and looks similar to that observed 
in B3\,V stars \citep{meyer98,ranade04}. Around 1.70\,$\mu m$ we find a weak feature, 
that could be a He\,I line, and the $K$ band spectrum is similar to the one observed in B3-5\,V 
type stars \citep{hanson96,hanson98}, which we adopt as spectral type for 4a.
	
Finally there are two stars in this group without a clear spectral classification. The first one is star 3a, with an irregular Brackett 
series and an asymmetric Br$\gamma$ line. We see the He\,I lines at 1.70\,$\mu m$ and at 2.06\,$\mu m$, but the last one 
is located at a wavelength with a problematic telluric correction. Because this spectrum is noisy and most of its Brackett series 
is within the noise level, we cannot clearly classify it. Both the presence of He\,I at 1.70 $\mu m$ and the absence of He\,II limit 
the star 3a spectral type to early B-type.

 The second star, 8a, presents the Brackett series in strong emission for both $H$ and $K$ bands. No extended emission 
was detected during the reduction process, indicating that the emission comes from the immediate vicinity of the star. Its spectrum 
shows Fe\.II at 1.64\,$\mu m$ in emission, which indicates that it is very young, probably still surrounded by circumstellar 
material, as described by \citet{puga10} for their star 26. Because of the complete emission Brackett series, the emission Fe\,II 
line at 1.64\,$\mu m$ and the colours of this star, we classify it as a young stellar object (YSO).

\subsubsection{A-, F-, and G-type dwarfs}
 The observation of this type of stars was expected as contamination, because the pseudocolour $Q_{IR}$ for OB and 
AF-type dwarfs is similar. The earliest star in this group, 8b, only presents in its spectrum a clear Brackett series, extended 
until the H\,I (4-18) line. No metallic lines are clearly distinguished in this spectrum, and we assign it the A5\,III-V 
spectral type.

 For F-type stars, the Brackett series begins to diminish, and Mg\,I lines at 1.57, 1.71, and 2.2\,$\mu m$ become stronger. 
In the star 9b spectrum the H\,I (4-11), (4-12), and (4-13) lines are present and similar to those observed in 
F3\,V \citep{meyer98} or F7\,V stars \citep{meyer98}. Even if Mg\,I is not clearly seen in the star 9b spectrum, 
we assign the F3-7\,V spectral type .

 For stars 6a (F6\,V), 11a (F6-8\,V), and 10a (F8\,V), the Brackett series does not
extend beyond the H\,I (4-13) line, their Br$\gamma$ lines are narrow and deep, and the Mg\,I lines in the $H$ band 
spectrum are similar to those seen in late F\,V stars \citep{meyer98}. 

The spectrum of star 5b presents late F-type features and a faint ${}^{12}$CO\,(2,0) band. The presence of this 
band indicates an early G-type for this star. The $H$ band spectrum is similar to an F6\,V star \citep{meyer98}, 
but the $K$ band spectrum is more similar to an F8.5\,V star, considering the Br$\gamma$ line's depth and the Mg\,I 
line at 2.28\,$\mu m$ \citep{meyer98}. Because of the faint presence of the CO band, we assign this star a spectral 
type between F6 and G1\,V.

 The spectrum of star 4b presents clear Mg\,I lines in $H$ and $K$ bands; its Mg\,I lines at 1.57 and 1.71\,$\mu m$ 
are broad but not as deep as expected for a G2\,V star. The $K$ band spectrum does not show the Br$\gamma$ line, 
discarding a luminosity class III for this object, and its ${}^{12}$CO\,(2,0) band fit the band from a G2-3\,V star.

 In the case of the star 7b, we observe that the Mg\,I lines are similar in depth and shape, to a 
G8\,V or K0\,V star. The $K$ band spectrum shows a shallow ${}^{12}$CO\,(2,0) band and 
a deep Ca\,I line at 2.27\,$\mu m$. Even if the deep of the CO band is similar to the K0\,V or 
K5\,V CO band, the depth of the Ca\,I line indicates an earlier spectral type for this object, between 
G8 and K0\,V.
 
\subsubsection{Giant stars}
The three stars from this group present late-type spectral lines that allow us to classify them as luminosity class III.
The spectra of stars 5a, 7a, and 6b have clear ${}^{12}$CO\,(3,0) bands in $H$, characteristic of 
luminosity classes I or III, and not observed in stars with luminosity class V. The equivalent widths for the 
${}^{12}$CO\,(2,0) band in $K$, between 2.294 ${\mu m}$ and 2.304 ${\mu m}$, are $\mathrm{{EW}_{6b}}=17.60$ \AA, 
$\mathrm{{EW}_{5a}}=4.23$\AA, and $\mathrm{{EW}_{7a}}=8.96 $\AA; these values fit better a luminosity 
class III for the three stars, according to the equivalent width and luminosity class relation given by \citet{davies07},  
and the spectral type adopted for these three stars.
 
The star 5a spectrum is similar to a G4III or a G7III spectra \citep{wallacehinkle97} in the $K$ band. In the $H$ band 
it is similar to a G6\,III \citep{meyer98}. We adopt the spectral type G4--7\,III for this star.    

 The CO depth in $H$ for star 7a is similar to that observed in the HR\,7949 spectrum (spectral type K0\,III, 
\citealt{meyer98}), and the CO band in $K$ fits the HR\,8694 (spectral type K0\,III, \citealt{wallacehinkle97}) 
or the HR\,6299 band (spectral type K2\,III, \citealt{wallacehinkle97}). For this star we adopt a 
spectral type K0--2\,III.
 
 Finally, for the star 6b we have adopted a M0--1\,III spectral type, based on its $K$ band spectrum, which is 
similar to the HR\,4069 and the HR\,7635 spectra (spectral types M0\,III, \citealt{wallacehinkle97}), and its $H$ band 
spectrum, similar to the HR\,4517 spectrum (spectral type M1\,III, \citealt{meyer98}).


\section{Discussion}\label{discusion_m04}

  \subsection{Distance estimate}

 From the stars observed spectroscopically, 19 out of 21 were selected following the pseudocolour $Q_{IR}$ 
criteria (including star 11a, which has $Q_{IR}=0.25$, slightly beyond the upper limit). The other two (i.e. 7a 
and 6b) were chosen because of their field position. From the 19 stars, 10 were classified as 
OB-type dwarfs.
 
 For all the stars but one (i.e. star 8a), we derive individual distances by comparing apparent and intrinsic magnitudes according 
 to the assigned spectral type. We assume visual magnitudes from \citet{cox00}, intrinsic infrared colours from \citet{tokunaga00},  
 and the extinction law from \citet{rieke89} with $R=3.09$ \citep{rieke85}. The choice of this extinction law was justified in Section 
 \ref{diagramasM04}. The $K_S$-filter extinction, expressed as

\begin{equation}
A_{K_S} = \frac{E_{J-K_S}}{1.514} 
\end{equation}
varies between 0.04 mag and 4.19 mag, equivalent to $A_V$ between 0.37 mag and 38.66 mag.

 Distance errors were dominated by the spectral type uncertainties, which we estimate by deriving the individual distance for the same 
star assuming $\pm 2$ spectral subtypes. The distance errors associated with photometric errors were generally less than 10\% of those 
derived from the spectral type uncertainty and therefore were not included in the error estimates. For stars with a range in spectral type 
(for example star 1a, B2--3\,V), we assume the individual distance estimate to be the mean between the 
distance associated with each spectral type defining the range. 
 
\begin{figure}
\centering
\includegraphics[width=7.0cm,angle=270]{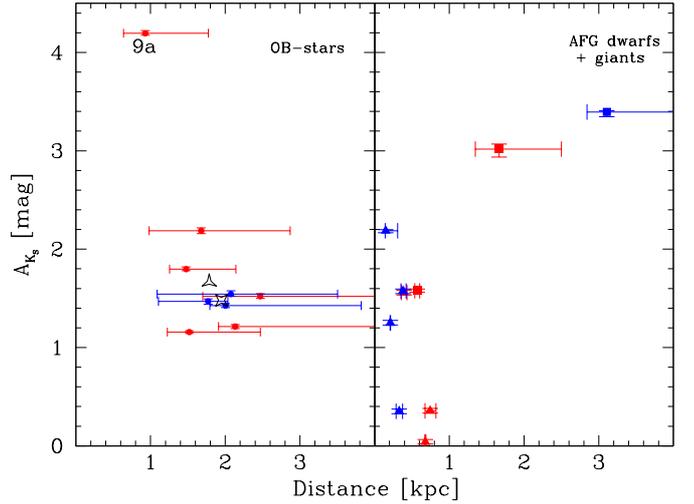}
       \caption{\textit{Left:} Individual distances and extinctions for OB-type stars. Red symbols show the parameters for region A stars, and blue 
       symbols for region B stars. Mean values of distance and extinction for regions A and B stars are marked with three- 
       and four-pointed black stars, respectively. \textit{Right:} Individual distances and extinctions for non-OB stars in regions A (red 
       symbols) and B (blue symbols). Estimates for spectral type A, F, and G dwarfs are shown with triangles, and giant stars estimates are shown with 
       squares.}
       \label{m04_ext_dist}
\end{figure}

  In Table \ref{msgms04_data_stars} we present the individual extinction and distance determination for the OB-type stars used in 
the cluster distance estimate. These estimates are also shown in Figure \ref{m04_ext_dist}, where we present regions A (red) and B 
(blue) individual distances and extinctions for OB (left) and non-OB (right) type stars. In the figure we show the mean distance 
value for regions A (three-pointed star) and B (four-pointed star), and we can observe that:
 
\begin{itemize}
\item{In region A, OB-type stars have a mean distance of $1.87^{+1.29}_{-0.77}$ kpc. For this determination, we have excluded 
stars 8a and 9a, both belonging to the nucleus of region A. Because star 8a has not been spectroscopically classified, the extinction 
and distance could not be estimated for this object. For star 9a, the spectroscopically observed star with the largest reddening, photometry 
is probably contaminated by local nebulosity, making the measured magnitude unreliable. 

 Even without an extinction or distance estimate for stars 8a and 9a, their positions in the CCDs indicate that 
these stars are not foreground objects.}
\item{In region B, a mean distance of $1.95^{+1.21}_{-0.76}$ kpc was derived from stars 1b, 2b, and, 3b individual distances.
This distance is in agreement within errors with the one derived for IRAS 18537+0753: 2.1 kpc \citep{plume92}, and the 
estimate derived from the observations of the methanol maser [HHG86] 185345.9+074916: 2.2 kpc \citep{valtts00}. 

 We conclude that stars 1b, 2b, and 3b are associated with the IRAS source IRAS 18537+0753, the methanol maser, 
and the massive star forming region Sh2-76 E.} 
\item{Dwarfs stars with spectral type A, F, or G have individual distances less than 1 kpc. Consequently we classify these objects 
as foreground dwarf stars from the galactic disc. 

 The exception in this group is star 7b, with a distance estimate of $d=5.44$ kpc. After reviewing the LIRIS images we  
 notice that this star is contaminated by a close companion, and their radial profiles are merged. Because of this contamination, 
 we exclude the star 7b from the analysis.}
\item{Giant stars have two types of extinction and distance values: the foreground star 5a, probably 
located in the same structure as A-, F-, and G-type stars; and the background giant stars 5a and 6b. 
In both cases these stars are part of the giant population of the galactic disc.}
\end{itemize}

 The OB-type stars from the cores in regions A and B are located at the same distance within errors. We thus conclude that they belong
 to two near-infrared nuclei of a single, young, massive, and embedded stellar population. Using the mean distance of regions A and B, we estimate 
 a distance of $1.90^{+1.28}_{-0.90}$ kpc for the Masgomas-4 association, with a mean extinction $A_{K}$=1.54$\pm$0.02 mag. At 
 this distance, the projected separation of regions A and B corresponds to a linear separation of about 2.4 pc. 

In the CCDs (Fig. \ref{m04_ccd}) we observe objects in the T Tauri region, below the black segmented line. The T Tauri
candidates are present in both regions A and B, but only in the case of region B we identify a group of objects with extra reddening (colours 
$(H-K_S)>2$, shown with red circles in the CCD of region B). These stars (or protostars) are concentrated around the region B nucleus, as can 
be seen in Figure \ref{m04_ks}, and their extra reddening can be explained by the presence of circumstellar accretion discs.

Because of the extra reddening of this group of stars, compared with the colours of the T Tauri objects, we classify them as 
Herbig Ae/Be candidates. According to \citet{subramaniam06}, Herbig Ae/Be objects are located to the right of the locus of 
the classic T Tauri objects in the CCDs. These objects are also younger, more embedded, and more massive than T Tauri
objects.

 Like Masgomas-1 \citep{ramirezalegria12}, Masgomas-4 is located in the direction of the RSG clusters. Our distance estimate 
discards a physical link of Masgomas-4 with the intersection between the close end of the Galactic bar and the Scutum-Centaurus 
arm base. The 1.90 kpc distance estimate places Masgomas-4 closer than the Scutum-Centaurus arm, probably in the Carina-Sagittarius 
arm.
 
\subsection{Mass and age estimates}

To estimate the cluster's present-day total mass, we follow a similar procedure to that described by \citet{ramirezalegria12}.
We fit a Kroupa \citep{kroupa01} initial mass function (IMF) to the cluster present-day mass function, and integrate
it from log $(M) = -1.0$ dex to 1.5 dex. Because there is not an evolved stellar population in Masgomas-4, 
we can assume that the difference between the initial and present mass function in the cluster is small.

 We fit the Kroupa IMF to the cluster mass function between $\sim$2.5 ${M}_{\odot}$ and $\sim$35 ${M}_{\odot}$. The 
mass function was derived from the luminosity function and the contribution from the galactic disc stellar population was 
corrected using the observed control field (from LIRIS images). The cluster and control field photometries were cut in 
$J<16.5$ mag, $H<15.5$ mag, and $K<14.5$ mag, to assure a completeness close to 1.0, following the same procedure 
described by \citet{ramirezalegria12}.

 To obtain the luminosity and mass functions, we project every star, following the reddening vector, to the main 
sequence located at 1.9 kpc. This sequence is defined by the colours and magnitudes given by \citet{cox00}, and is 
expressed analytically by two lines: one from the spectral type O9\,V to A0\,V, and the second one from A0\,V to G0\,V, 
corresponding to the last spectral type to the limit magnitude chosen for completeness.

 To derive the mass function we convert the $K_S$ magnitudes to solar masses, using the values given by 
\citet{cox00} for stars later than B0\,V, and \citet{martins05} for stars earlier than this. For magnitudes that were  
between values in the catalogue, we interpolate between the two closest values.

\begin{figure}
\centering
\includegraphics[width=3.5in]{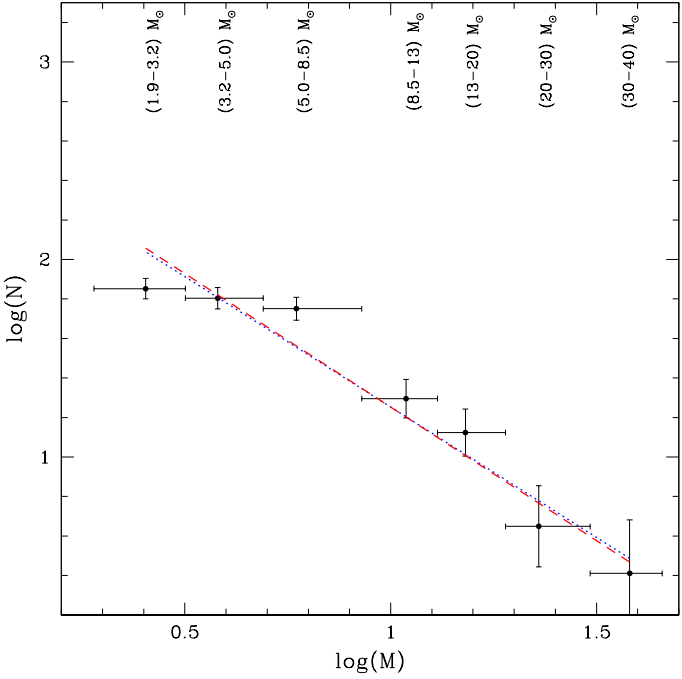}
            \caption{Present-day mass function for Masgomas-4. Points show the central position in the mass ranges 
            indicated above them, and the segmented red line corresponds to the Kroupa IMF fitted to the data. 
            As a comparison, we show the least-squares fitting (one degree function) to our data with a blue dotted line.}
       \label{histograms_m04}
    \end{figure}

 After subtracting the Masgomas-4 field and the control field mass function, we obtain the cluster present-day 
mass function, shown in Figure \ref{histograms_m04}. We fit a Kroupa IMF and integrate in the mass range 
0.10-31 ${M}_{\odot}$ (equivalent to log $(M) = -1.0$ dex to 1.5 dex), estimating a total cluster mass (lower limit) of 
$(2.19\pm0.31)\cdot10^3 {M}_{\odot}$. The Kroupa IMF fitting is also compared in the figure with a least-squares  
fitting to the data. Both fits are similar within data errors.

 Cluster age can be limited by the presence of an O9\,V star. This gives an upper limit of 10 Myr that could be 
reduced by the confirmation of earlier stars in the centre of region A, for example star 8a. 
This age limit of 10 Myr is also dependent on the rotation speed assumed by the stellar evolutionary models \citep{brott11}.

 The presence of T Tauri and Herbig Ae/Be candidates also gives information about the cluster age. According 
to \citet{hernandez07} in young clusters (age $\sim3$ Myr), 30\% of the T Tauri objects and 15\% of the Herbig 
Ae/Be objects lose their circumstellar discs. The survivor discs around Herbig Ae/Be objects will be lost in 
less than 10 Myr. 

 Although there is a difference because of the presence of Herbig Ae/Be candidates in region B, which are absent 
 in region A, both cores display signs of on-going star formation (IRAS sources and masers). The presence of Herbig
 Ae/Be candidates also limits the age of the recent burst in the region B core to 5 Myr.
 
\citet{lada03} indicate that it is very unlikely to find embedded clusters older than 5 Myr. Because part of the stellar 
population of Masgomas-4 is still deeply embedded and the whole cluster is surrounded by an H\,II cloud, we  
adopt this value as a new upper limit  for the cluster age, limiting the cluster age to 5 Myr.


\section{Conclusions}\label{conclusiones}

 We present the observations, physical estimates, and analysis for the young, double-core, and obscured cluster 
 Masgomas-4. The near-infrared observations, individual extinctions, and distance estimates for the OB-type 
 star population are in agreement with a single association, with two active 
 regions of stellar formation. The individual distances estimated for regions A and B point to a
 single distance of $1.90^{+1.28}_{-0.90}$ kpc, and this distance indicates that the cluster is associated with 
the IRAS source IRAS 18537+0753, the methanol maser [HHG86] 185345.9+074916, and the massive star 
forming region Sh2-76 E. The Algol variable star \object{V 1665 Aql} is discarded as part of the Masgomas-4 
stellar population.

 We observe in both regions A and B massive stars, and spectroscopic evidence of young or forming stars, 
like central stars 8a (emission Brackett series and Fe\,II at 1.64 $\mu m$ in emission) and 9a (Fe\,II at 1.64 $\mu m$ in 
emission). Regions A and B presents T Tauri candidates, according to their CCDs. In the case of 
region B, we also detect a clearly concentrated population of Herbig Ae/Be candidates around Sh2-76E. Its 
exclusive presence around the region B core could indicate a different star formation history between regions A and B
(due to initial mass or age) and a slightly younger population in the region B core.

The mass estimate, derived from integrating the Kroupa initial mass function fitted to our data, gives a lower limit 
for the cluster total mass of $(2.19\pm0.31)\cdot10^3$ $M_{\odot}$. The presence of T Tauri and Herbig Ae/Be 
candidates and the surrounding gas observed in mid-infrared images imply that Masgomas-4 is still a young and 
active object in terms of star formation. The very young obscured population and the surrounding gas allow us to 
limit the cluster age to less than 5 Myr.
 
 
\begin{acknowledgements}

S.R.A. was supported by the GEMINI-CONICYT project number 32110005, the FONDECYT project No. 3140605, 
and the investigation project Massive Stars in Galactic Obscured Massive Clusters. Part of this work was supported 
by the Science and Technology Ministry of the Kingdom of Spain (grants AYA2010-21697-C05-04 and AYA2012-39364-C02-01), 
the Gobierno de Canarias (PID2010119), and by the Chilean Ministry for the Economy, Development, and Tourism's Programa 
Inicativa Cient\'ifica Milenio through grant IC 12009, awarded to The Millennium Institute of Astrophysics (MAS).
 
 The William Herschel Telescope is operated on the island of La Palma by the Isaac Newton Group in the 
 Spanish Observatorio del Roque de los Muchachos of the Instituto de Astrof\'isica de Canarias. This publication 
 makes use of data products from the Two Micron All Sky Survey, which is a joint project of the University of 
 Massachusetts and the Infrared Processing and Analysis Center/California Institute of Technology, funded by the 
 National Aeronautics and Space Administration and the National Science Foundation. Average asymmetrical 
 errors were estimated using the Java applet available in http://www.slac.stanford.edu/$\sim$barlow/statistics.html
 \citep{barlow04}.

\end{acknowledgements}

\bibpunct{(}{)}{;}{a}{}{,} 
\bibliographystyle{aa} 
\bibliography{biblio}

\listofobjects

\end{document}